\documentclass[twocolumn,showpacs,preprintnumbers,amsmath,amssymb]{revtex4}

\usepackage{color}
\usepackage{graphicx}
\usepackage{dcolumn}
\usepackage{bm}

\newcommand{\eq}[1]{Eq.~(\ref{#1})}
\newcommand{\fig}[1]{Fig.~\ref{#1}} 

\newcommand{\dd}{\ensuremath{D\!-\!\Dbar{}\,}}
\newcommand{\kk}{\ensuremath{K\!-\!\Kbar{}\,}}

\newcommand{\ddm}{\dd\ mixing}

\newcommand{\Dbar}{\,\overline{\!D}}
\newcommand{\Kbar}{\,\overline{\!K}}

\begin{document}

\preprint{TTP10-17}

\title{Do squarks have to be degenerate?\\ Constraining the mass splitting with \kk and \dd mixing}
%
\author{Andreas Crivellin and Momchil Davidkov}
\affiliation{Institut f\"ur Theoretische Teilchenphysik\\
               Karlsruhe Institute of Technology, 
               Universit\"at Karlsruhe, \\ D-76128 Karlsruhe, Germany}%
\date{March 6, 2010}
\begin{abstract}
We study the constraints on the mass-splitting of the first two generations of left-handed squarks obtained from $\Delta M_K$, $\epsilon_K$ and \ddm. The different contributions from gluino, neutralino and chargino diagrams are examined in detail, concluding that it is not justified to neglect electroweak gaugino diagrams if the squark mass matrices contain flavor nondiagonal LL elements. We find that the constraints on the mass-splitting are very strong for light gluino masses. However, if the gluino is heavier than the squarks the constraints on the mass-splitting are much weaker. There are even large regions in parameter space where the different NP contributions cancel each other, leaving the mass-splitting nearly unconstrained.
\end{abstract}
\maketitle

\section{\label{sec:level1}Introduction}

Already in the early stages of minimal supersymmetric standard model (MSSM) analyses it was immediately noted, that a super GIM mechanism is needed in order to satisfy the bounds from flavor changing neutral currents (FCNCs) \cite{Dimopoulos:1981zb}. Therefore, the mass matrix of the left-handed squarks should be (at least approximately) proportional to the unit matrix, since otherwise flavor off-diagonal entries arise inevitably either in the up or in the down sector due to the SU(2) relation between the left-handed squark mass terms. 
The idea that nondegenerate squarks can still satisfy the FCNC constraints (K and D mixing) was first discussed in Ref.~\cite{Nir:1993mx} (an updated analysis can be found in Ref.~\cite{Nir:2002ah}) in the context of Abelian flavor symmetries \cite{Dudas:1995eq,Agashe:2003rj}. In the meantime, there have been a lot of significant improvements both on the theoretical and on the experimental side: The mass difference in the D system was measured and the decay constants and bag factors were calculated to a high precision using lattice methods. A recent analysis of the constraints put on NP by Kaon and D mixing can be found in \cite{Blum:2009sk}. In all MSSM analyses the main focus has been on the gluino contributions, while the chargino and neutralino contributions were usually neglected claiming that they are suppressed by a factor of $g_2^4/g_s^4$  \cite{Nir:1993mx, Hagelin:1992tc, Gabbiani:1996hi, Ciuchini:1998ix, Ciuchini:2007cw, Blum:2009sk, Gedalia:2009kh}. However, it is no longer a good approximation to consider only the gluino contributions in the presence of off-diagonal elements in the LL block of the squark mass matrices because the winos couple to left-handed squarks with $g_2$. In addition, the gluino contributions suffer from cancellations between the crossed and uncrossed box-diagrams, especially if the gluino is heavier than the squarks. Therefore, the neutralino and chargino contributions can even be dominant if $M_2$ is light and the gluino is heavier than the squarks. This situation can occur in GUT-motivated scenarios in which the relation $M_2\approx m_{\tilde{g}} \alpha_2/\alpha_3$ holds. Therefore, we want to update the evaluation of the constraints from K and D mixing with focus on the mass splitting between the first two squark generations taking into account the weak contributions as well.

The squark spectrum is a hot topic concerning bench-mark scenarios for the LHC. It is commonly assumed that the squarks are degenerate at some high scale and that nondegeneracies are introduced via the renormalization group \cite{sps,gamma}. In such scenarios, the nondegeneracies are proportional to Yukawa couplings and therefore only sizable for the third generation. However, flavor-off-diagonal entries in the squark mass matrix can also lead to nondegenerate squarks which can have an interesting impact on the expected decay and production rate of squarks \cite{Hurth:2009ke}. In principle, there remains the possibility that squarks have already different masses at some high scale. The question which we want to clarify in this article is which regions in parameter space with nondegenerate squarks are compatible with \dd\ and \kk\ mixing. We are going to discuss this issue in Sec. III after reviewing \kk\ mixing  and  \dd\ mixing in Sec. II. Finally we conclude in Sec. IV.

\section{Meson mixing between the first two generations}

Measurements of flavor-changing neutral current (FCNC) processes put strong constraints on new physics at the TeV scale and provide a important guide for model building. In particular \dd\ and \kk\ mixing strongly constrain transitions between the first two generations and combining both is especially powerful to place bounds on new physics \cite{Blum:2009sk}. 
In the down sector FCNCs between the first two generations are probed by the neutral Kaon system, the first observed example of meson-antimeson mixing. Kaon mixing was already discovered in the early 1950s and the CP violation was established in 1964. The up to date experimental values for the mass difference and the CP violating quantity $\epsilon_K$ are \cite{PDG}:
\begin{eqnarray}
\Delta m_K/m_K=(7.01\pm0.01)\times10^{-15}\nonumber\\
\epsilon_K=(2.23\pm0.01)\times10^{-3}
\label{Kmixing}
\end{eqnarray}
However, still today, in the age of the B-factories, the long known neutral Kaon system still provides powerful constraints on the flavor structure of any NP model. As we see from \eq{Kmixing} both the mass difference and the size of the indirect CP violation are tiny and the numbers are in agreement with the standard model (SM) prediction: The SM contribution to the mass difference is small due to a rather precise GIM suppression (the top contribution is suppressed by small CKM elements) and also the CP asymmetry is strongly suppressed because CP violation necessarily involves the tiny CKM combination $V_{td}V_{ts}^*$ related to the third fermion generation. Therefore, Kaon mixing puts very strong bounds on NP scenarios like the MSSM. According to the analysis of Ref.~\cite{Ciuchini:2000de} the allowed range in the $C_{M_K}-C_{\epsilon_K}$ plane is rather limited. At 95\% confidence level on can roughly expect the NP contribution to the mass difference $\Delta M_K$ to be at most of the order of the SM contribution. The NP contribution to $\epsilon_K$ is even more restricted. The gluino contribution to \kk\ mixing was in the focus of many analyses \cite{Dimopoulos:1981zb, Nir:1993mx, Hagelin:1992tc, Gabbiani:1996hi}. An complete study of the gluino contributions, taking into account the NLO evolution of the Wilson coefficients was done in Ref. \cite{Ciuchini:1998ix}. (In a recent analysis \cite{Crivellin:2009ar} we pointed out the importance of (N)NLO chirally enhanced corrections in the presence of nondegenerate squark masses for the constraints on $\delta^{d\;LR}_{12,21}$ obtained from Kaon mixing and Ref.~\cite{Virto:2009wm} calculated the full NLO matching for the gluino contributions. The NLO corrections were  calculated previously in Ref.~\cite{Ciuchini:2006dw}, however this analysis was not sensitive to chirally enhanced corrections since the squarks were assumed to be degenerate and the chirally enhanced corrections 
drop out in this case.) However, neither of these articles considered the electroweak contributions. Only Ref.~\cite{Khalil:2001wr} calculated the chargino contributions but the gluino and neutralino contribution were neglected in this article and the SU(2) relation connecting the up and down squark mass matrices was not used. We return to this point Sec. III.

In the up sector FCNCs are probed by \dd\ mixing. In contrast to the well-established Kaon mixing, it was only discovered recently in 2007 by the BABAR \cite{Aubert:2007wf} and BELLE \cite{Staric:2007dt,Abe:2007rd} collaborations. The current experimental values are \cite{Barberio:2008fa}: 
\begin{eqnarray}
\Delta m_D/m_D=(8.6\pm2.1)\times10^{-15}\nonumber\\
A_\Gamma=(1.2\pm2.5)\times10^{-3}
\end{eqnarray}
Short-distance SM effects are strongly CKM suppressed and the long-distance contributions can only be estimated. Therefore, conservative estimates assume for the SM contribution a range up to the absolute measured value of the mass difference. However, due to the small measured mass difference D mixing still limits NP contributions in a stringent way. Furthermore, a CP phase in the neutral D system can directly be attributed to NP. A first analysis (also including the implications for the MSSM) was done shortly after the experimental discovery \cite{Ciuchini:2007cw} and a recent update can be found in Ref.~\cite{Gedalia:2009kh}. However, these studies did not consider the electroweak contributions. 

In summary, \dd\ and \kk\ mixing restrict FCNC interactions between the first two generations in a stringent way and one should expect the NP contributions to the mass difference to be smaller than the experimental value \cite{Blum:2009sk}:
\begin{equation}
\Delta m^{\rm{NP}}_{D,K}\leq \Delta m^{\rm{exp}}_{D,K}
\label{DeltaM}
\end{equation}
CP violation associated with new physics is even more restricted, especially in the d sector:
\begin{equation}
\epsilon^{\rm{NP}}_{K}\leq 0.6\epsilon^{\rm{exp}}_{K}
\label{epsilon}
\end{equation}
Equations (\ref{DeltaM}) and (\ref{epsilon}) summarize in a concise way the allowed range for NP and we will use them to constrain the NP contributions to K and D mixing in Sec.~\ref{DK}.

\section{Constraints on the mass splitting from Kaon mixing and D mixing.\label{DK}}

In this section we want to discuss the constraints on the mass splitting between the first two generations of left-handed squark. Because of the $SU(2)$ relation between the left-handed up and down squark mass matrices, $M^{2}_{\tilde{u}}=V_{CKM}^{\dagger}M^{2}_{\tilde{q}}V_{CKM}$, in the super-CKM basis, these mass matrices are not independent. The only way to avoid flavor off-diagonal mass insertions in the up and in the down sector simultaneously is to chose $M^{2}_{\tilde{q}}$ proportional to the unit matrix. This is realized in the naive minimal flavor violating MSSM. 
In a more general definition of MFV \cite{Isidori} flavor-violation due to NP is postulated to stem solely from the Yukawa sector, resulting in FCNC transitions (which can now also be mediated by gluinos and neutralions) proportional to products of CKM elements and Yukawa couplings. Therefore, such scenarios allow only sizable deviations from degeneracy with respect to the third generation. However, even though nondegeneracies with the third generation induce additional CP violation associated with $V_{ub}$ we find that this mass splitting effectively cannot be constrained. This finding is in agreement with Ref.~\cite{Gedalia:2010zs} 
A bit more general notion of MFV could be defined by stating that all flavor change should be induced by CKM elements. This definition would also cover the case with a diagonal squark mass matrix in one sector (either the up or the down sector) but with off-diagonal elements, introduced by the $SU(2)$ relation, in the other sector. This setup corresponds to an exact alignment of the squark mass term $m_{\tilde{q}}^2$ with the product of Yukawa matrices $Y_u^{\dagger}Y_u$ (or with $Y_d^{\dagger}Y_d$ in the case of a diagonal down squark mass matrix).

The obvious way how off-diagonal elements of the squark mass matrices enter meson mixing is via squark-gluino diagrams. These contributions are commonly expected to be dominant since they involve the strong coupling constant. Also in our case under study, with flavor-violating LL elements, the gluino diagrams were assumed to be the most important SUSY contributions to the Wilson coefficient $C_1$ of the $\Delta \rm{F}=2$ effective Hamiltonian $H_{eff}^{\Delta F=2}=\sum_{i=1}^5 C_i O_i+\sum_{i=1}^3 \tilde{C}_i \tilde{O}_i$ \cite{Nir:1993mx, Hagelin:1992tc, Gabbiani:1996hi, Ciuchini:1998ix, Ciuchini:2007cw, Blum:2009sk, Gedalia:2009kh}:
\begin{widetext}
\begin{equation}
C_1^{\tilde g\tilde g}  =  - \frac{{g_s^4 }}{{16\pi ^2 }}\sum\limits_{s,t = 1}^6 {\left[ {\frac{{11}}{{36}}D_2 \left( {m_{\tilde q_s }^2 ,m_{\tilde q_t }^2 ,m_{\tilde g}^2 ,m_{\tilde g}^2 } \right) + \frac{1}{9}m_{\tilde g}^2 D_0 \left( {m_{\tilde q_s }^2 ,m_{\tilde q_t }^2 ,m_{\tilde g}^2 ,m_{\tilde g}^2 } \right)} \right]V_{s\;12}^{q\;LL} V_{t\;12}^{q\;LL} } 
\label{C1gg}
\end{equation}
\end{widetext}
Our conventions for the loop-functions and the matrices in flavor space $V_{s\;12}^{q\;LL}$ are given in the appendix of Ref.~\cite{Crivellin:2008mq}. However, if we have flavor-changing LL elements it is no longer possible to concentrate on the gluino contributions for four reasons:
\begin{itemize}
	\item The gluino contributions suffer from cancellations between the boxes with crossed and uncrossed gluino lines corresponding to the two terms in the square brackets in \eq{C1gg}. The crossed box diagrams occur since the gluino is a majorana particle. This cancellation occurs approximately in the region where $m_{\tilde{g}}\approx 1.5 \,m_{\tilde{q}}$.
	\item In the SU(2) limit with unbroken SUSY the winos couple directly to left-handed particles with the weak coupling constant $g_2$. Therefore, flavor-changing LL elements can contribute without involving small left-right or gaugino mixing angles.
	\item Since charginos are Dirac fermions, there are no cancellations between different diagrams at the one-loop order. 
	\item The wino mass $M_2$ is often assumed to be much lighter than the gluino mass. In most GUT models the relation $M_2\approx m_{\tilde{g}}\alpha_{2}/\alpha_{3}$ holds. Since the loop function is always dominated by the heaviest mass, one can expect large chargino and neutralino contributions if the squarks masses are similar to the lighter chargino masses.
\end{itemize}
Therefore, we have to take into account the weak (and the mixed weak-strong) contributions to $C_1$: 
\begin{widetext}
\begin{align}
C_1^{\tilde \chi ^0 \tilde \chi ^0 }  &=  - \frac{1}{128\pi ^2 }\frac{g_2^4 }{4}\sum\limits_{s,t = 1}^6 \left( D_2 \left( m_{\tilde q_s }^2 ,m_{\tilde q_t }^2 ,M_2^2 ,M_2^2  \right) + 2 M_2^2 D_0 \left( m_{\tilde q_s }^2 ,m_{\tilde{q}_t }^2 ,M_2^2 ,M_2^2  \right) \right)V_{s\;12}^{q\;LL} V_{t\;12}^{q\;LL} & \nonumber
\\
C_1^{\tilde g\tilde \chi ^0 }  &=  - \frac{1}{16\pi ^2 }\frac{g_s^2 g_2^2 }{2}\sum\limits_{s,t = 1}^6 \left( \frac{1}{6}D_2 \left( m_{\tilde q_s }^2 ,m_{\tilde q_t }^2 ,m_{\tilde g}^2 ,M_2^2  \right) + \frac{1}{3}m_{\tilde g} M_2 D_0 \left( m_{\tilde q_s }^2 ,m_{\tilde q_t }^2 ,m_{\tilde g}^2 ,M_2^2  \right) \right)V_{s\;12}^{q\;LL} V_{t\;12}^{q\;LL} & \label{C1ew}
\\
C_1^{\tilde \chi ^ +  \tilde \chi ^ +  } &=  - \frac{g_2^4}{128\pi ^2 }\sum\limits_{s,t = 1}^6 D_2 \left( m_{\tilde q_s }^2 ,m_{\tilde q_t }^2 ,M_2^2 ,M_2^2  \right)V_{s\;12}^{q\;LL} V_{t\;12}^{q\;LL}&  \nonumber 
\end{align}
\end{widetext}
In \eq{C1ew} we have set all Yukawa couplings to zero and neglected small chargino and neutralino mixing. Because of the small Yukawa couplings of the first two generations and the suppressed bino-wino mixing the only sizable contribution of both the gluino and the electroweak diagrams is to the same operator $O_1=\bar{s}\gamma^{\mu}P_L d\otimes \bar{s}\gamma_{\mu}P_L d$ as the SM contribution. Note that in all contribution the same combination of mixing matrices enters, since the CKM matrices in the chargino vertex cancels with the ones in the squark mass matrix. 
Reference~\cite{Altmannshofer:2007cs} calculated all Wilson coefficients contributing to $\Delta F=2$ processes in the MSSM and Ref.~\cite{Altmannshofer:2009ne} included also the chargino and neutralino contributions into their numerical analysis. However, the main focus of Ref. \cite{Altmannshofer:2009ne} is not on the mass-splitting between the first two squark generations and the importance of the different contributions is not apparent from the scatter plots used in their analysis.

\begin{figure}
\includegraphics[width=0.5\textwidth]{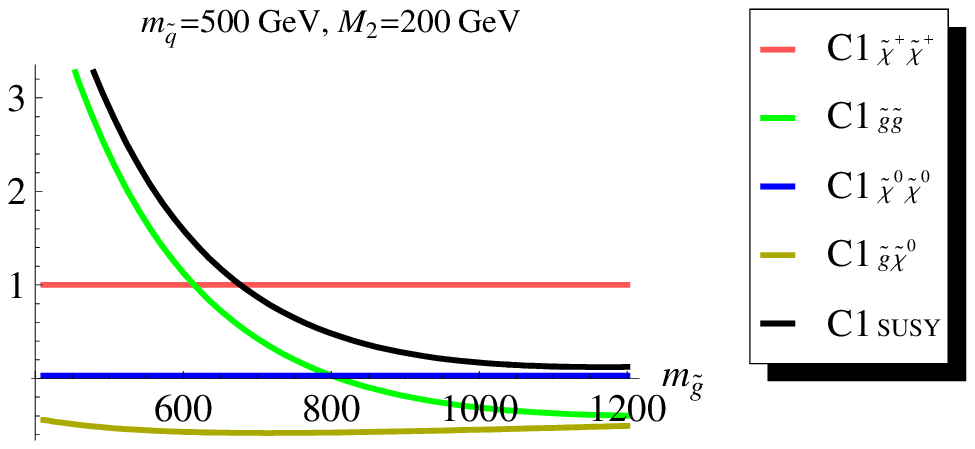}
\includegraphics[width=0.5\textwidth]{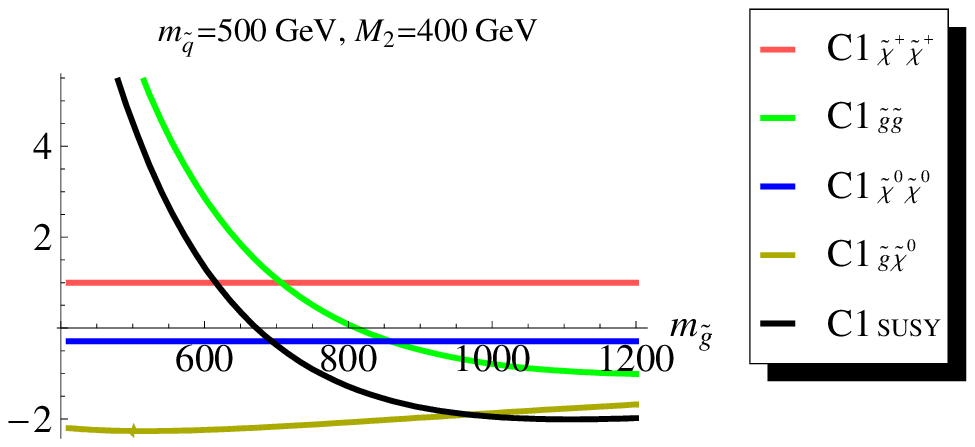}
\includegraphics[width=0.5\textwidth]{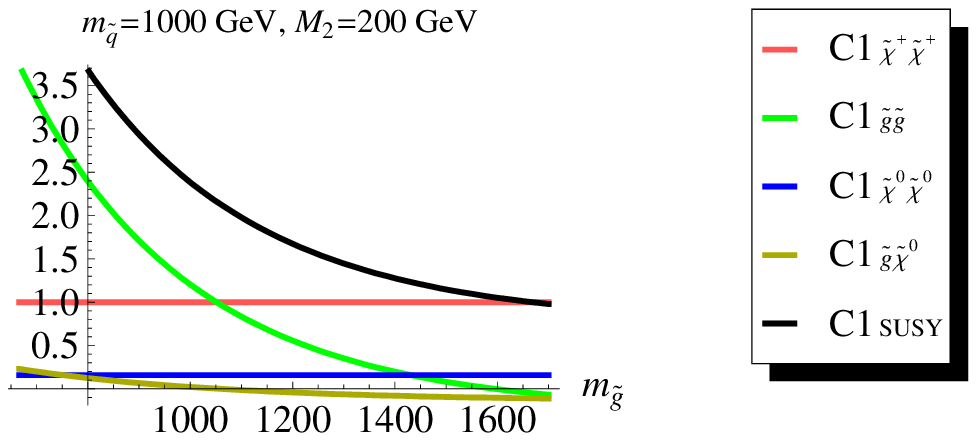}
\includegraphics[width=0.5\textwidth]{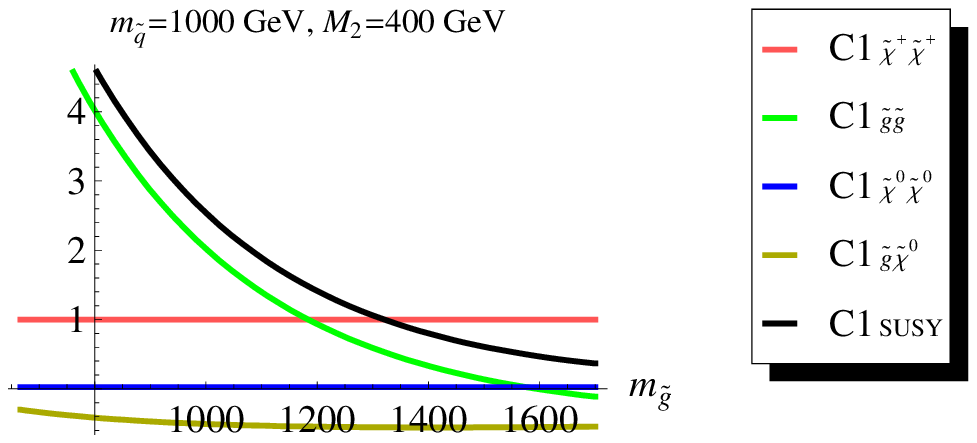}
\caption{Size of the real part of Wilson coefficients [see Eqs. (\ref{C1gg}) and (\ref{C1ew})] contributing to \dd\ or \kk\ mixing normalized to the chargino contribution as a function of $m_{\tilde{g}}$ for different values of $m_{\tilde{q}}$ and $M_2$ assuming a small nonzero (real) off-diagonal element $\delta^{q\;LL}_{12}$. $C_{1\rm{SUSY}}$ is the sum of all Wilson coefficients contributing in addition to the SM one. The relative size of the coefficients remains unchanged also in the case of complex elements $\delta^{q\;LL}_{12}$.}\label{C1}
\end{figure}

In \fig{C1} we show the size of the different contributions to $C_1$ as a function of the gluino mass. We have normalized all coefficients to $C_1^{\tilde \chi ^ +  \tilde \chi ^ + }$ since only one box diagram contributes to it and therefore the coefficient depends only on one loop-function which is strictly negative. Note that for heavy gluino masses always the chargino and in some cases the mixed gluino-neutralino contribution are dominant.

As stated before, SU(2) symmetry links a mass splitting in the up (down) sector to flavor-changing LL elements in the down (up) sector. So, if one assumes a "next-to minimal" setup in which one mass matrix is diagonal, one has to specify if this is the up or the down squark mass matrix. If the down (up) squark mass matrix is diagonal, which implies that it is aligned to $Y_d^{\dagger}Y_d$ ($Y_u^{\dagger}Y_u$), one has contributions to \dd\ (\kk) mixing.
Assuming a diagonal up-squark (down-squark) mass matrix, the regions in the $m_{\tilde{u}_1}$-$m_{\tilde{g}}$ plane compatible with \kk\ mixing (\dd\ mixing) are shown in Figs. (\ref{Constraints}) and (\ref{mass-splitting}). Note that there are large regions in parameter space with nondegenerate squark still allowed by \kk\ (\dd) mixing due to the cancellations between the different contributions shown in \fig{C1}. However, departing from an exact alignment with either $Y_u^{\dagger}Y_u$ or $Y_d^{\dagger}Y_d$ there are points in parameter space which allow for an even larger mass splitting \cite{Blum:2009sk} due to an additional off-diagonal element in the squark mass matrix. If this element is real one can choose an appropriate value which maximizes the allowed mass splitting \footnote{We thank Gilad Perez for bringing this to our attention.}. Nevertheless, this additional off-diagonal element now present in both sectors due to the SU(2) relation could also carry a phase additional to the CKM matrix. If this phase is maximal one obtains the minimally allowed range for the mass splitting due to the severe constraint from $\epsilon_K$. These minimally and maximally allowed regions for the mass splittings are also shown in Figs. (\ref{Constraints}) and (\ref{mass-splitting}).

We have seen that due to the cancellations between the different diagrams contributing to \dd\ and \kk\ mixing there are large allowed regions in parameter space where the squarks are not degenerate (a mass splitting of 100\% and more is well possible). This has also interesting consequences for the LHC: While most benchmark scenarios assume degenerate squark masses \cite{sps, gamma} nondegenerate masses can have interesting consequences on the branching ratios \cite{Hurth:2009ke}. The conclusion we can draw from Figs. (\ref{Constraints}) and (\ref{mass-splitting}) is that there are regions in parameter space, allowed by \kk and \dd~mixing, with very different masses for the first two squark generations. Therefore, FCNC processes alone do not require the soft-SUSY breaking parameter $M^{2}_{\tilde{q}}$ to be proportional to the unit matrix at some high scale. This implicates that there is more allowed parameter space for models with Abelian flavor symmetries than without the inclusion of the electroweak contributions to \dd\ and \kk\ mixing.

\section{Conclusions}

In this article we have examined the constraints on the mass splitting between the first two generations of left-handed squarks from \kk\ and \dd\ mixing by considering the gluino and the electroweak contribution.
\begin{widetext}
\newpage
\begin{figure}[h!]
\includegraphics[width=0.38\textwidth]{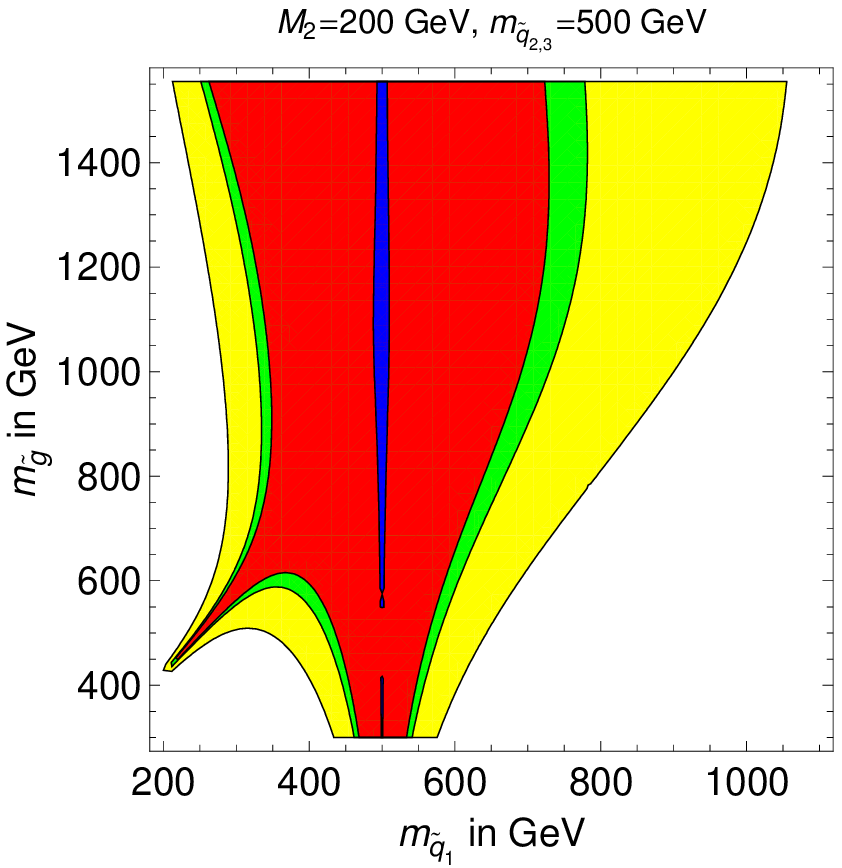}
\includegraphics[width=0.38\textwidth]{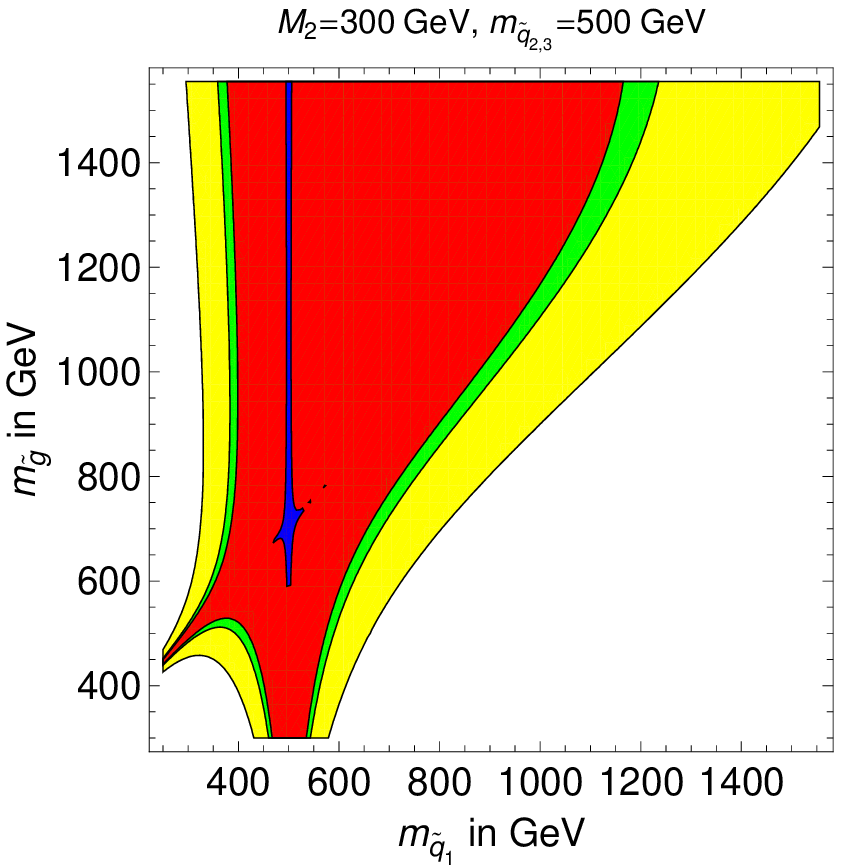}
\includegraphics[width=0.38\textwidth]{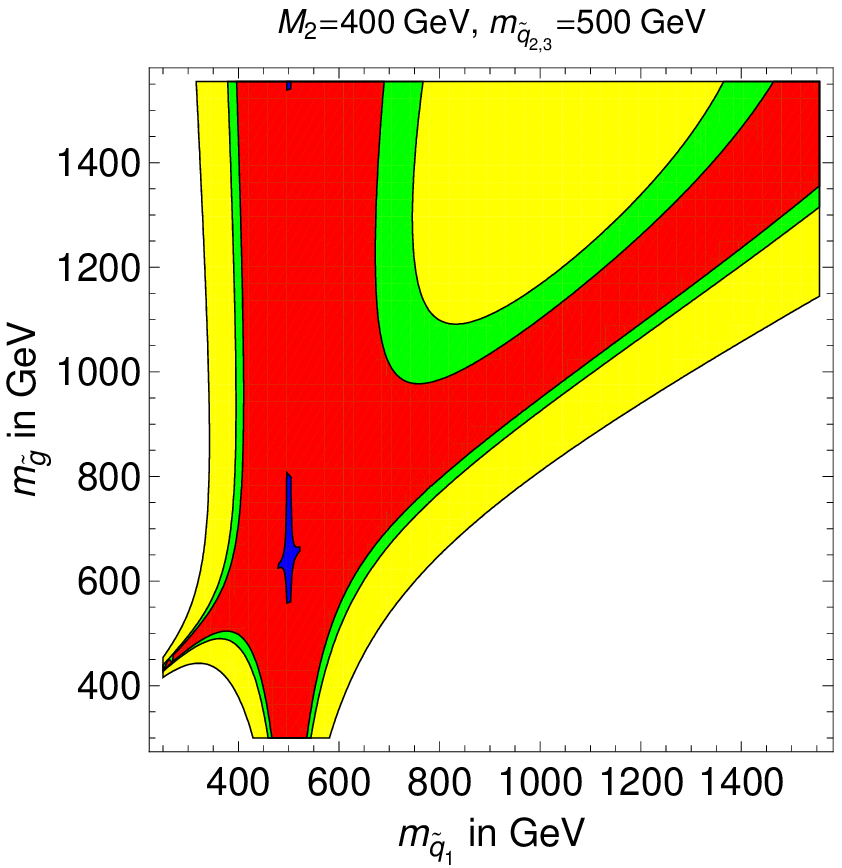}
\includegraphics[width=0.38\textwidth]{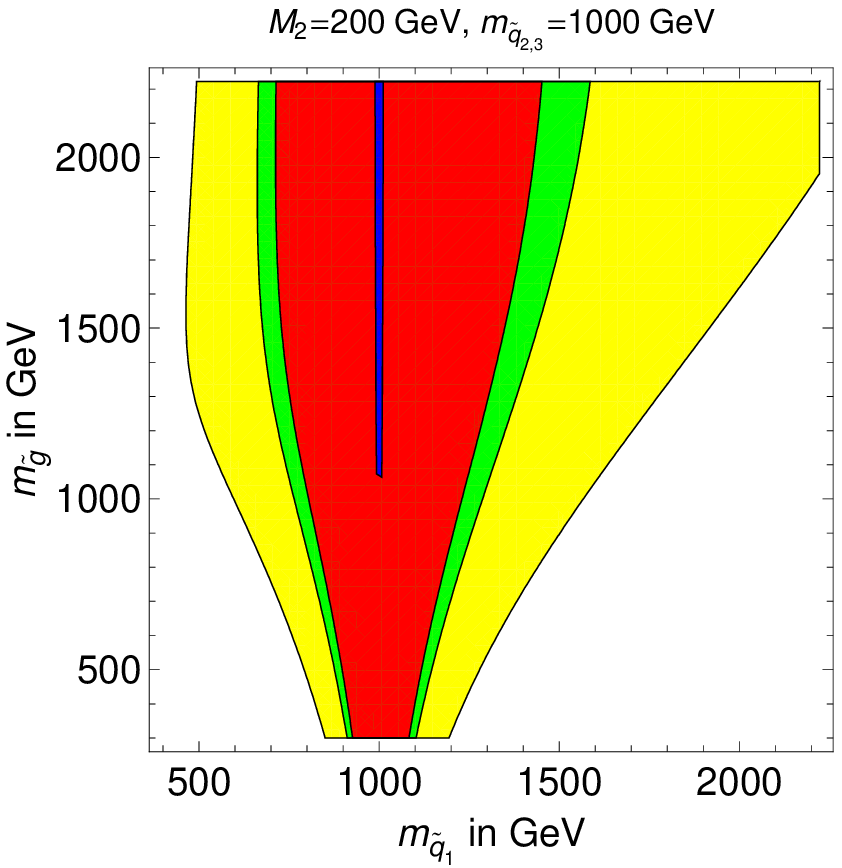}
\includegraphics[width=0.38\textwidth]{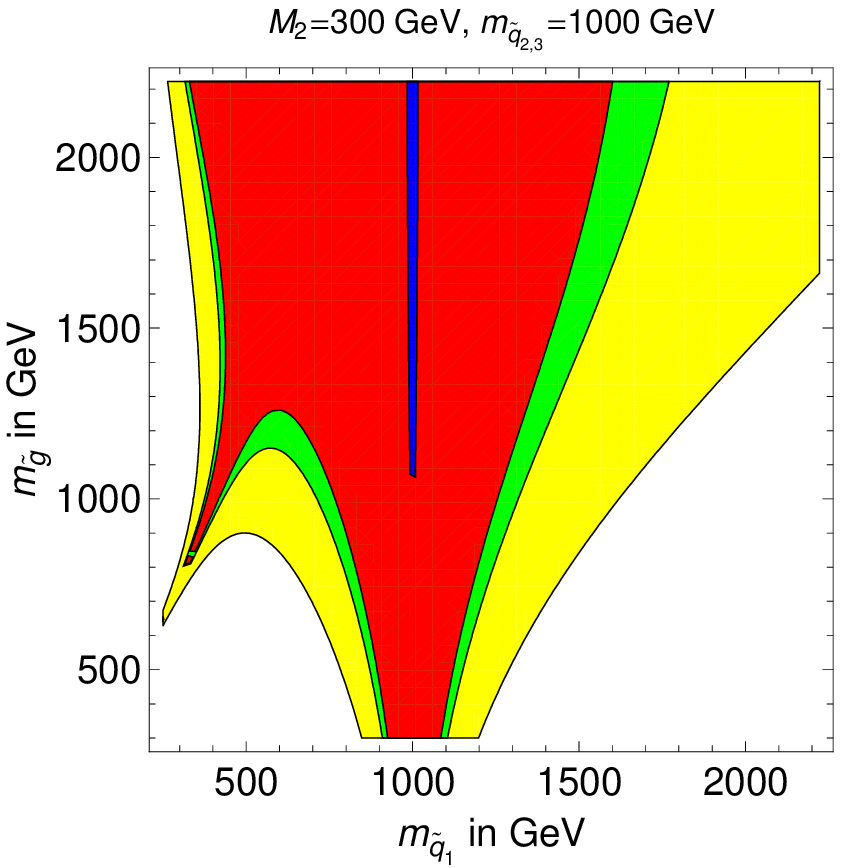}
\includegraphics[width=0.38\textwidth]{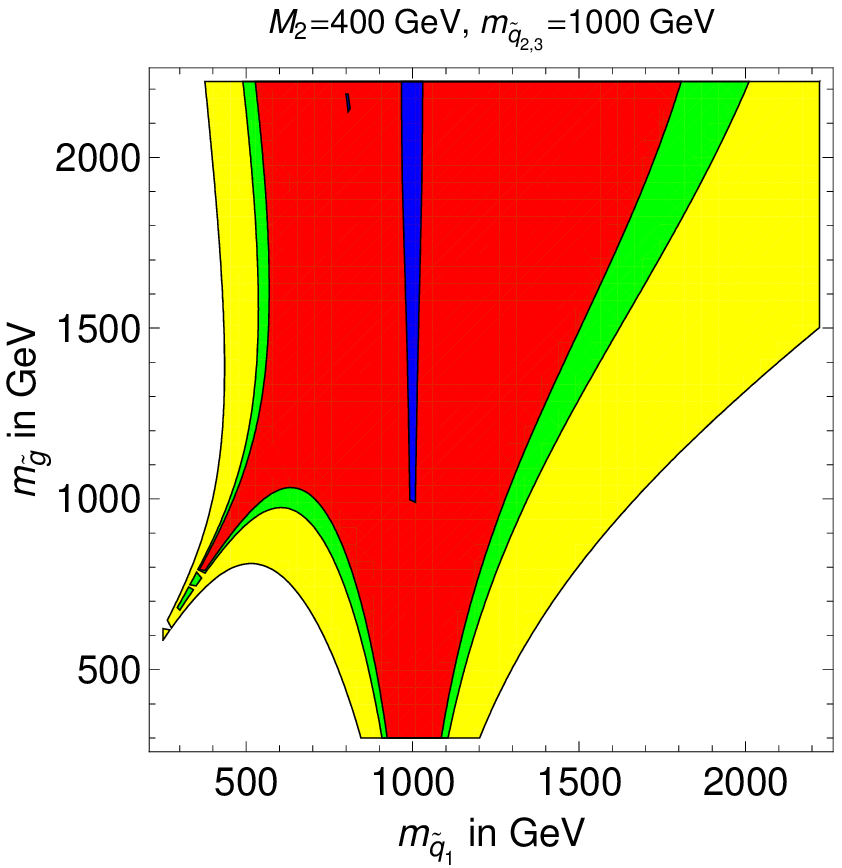}
\caption{Allowed regions according to Eqs. (\ref{DeltaM}) and (\ref{epsilon}) in the $m_{\tilde{q}_1}-m_{\tilde{g}}$ plane with $m_{\tilde{q}_{2,3}}=500 \,\rm{GeV}$ and $m_{\tilde{q}_{2,3}}=1000 \,\rm{GeV}$ for different values of $M_2$. Yellow (lightest) corresponds to the maximally allowed mass splitting assuming an intermediate alignment of $m^2_{\tilde{q}}$ with $Y_u^{\dagger}Y_u$ and $Y_d^{\dagger}Y_d$ \cite{Blum:2009sk}. The green (red) region is the allowed range assuming an diagonal up (down) squark mass matrix. The blue (darkest) area is the minimal region allowed for the mass splitting between the left-handed squarks, which corresponds to a scenario with equal diagonal entries in the down squark mass matrix but with an off-diagonal element carrying a maximal phase.}\label{Constraints}
\end{figure}
\begin{figure}
\includegraphics[width=0.38\textwidth]{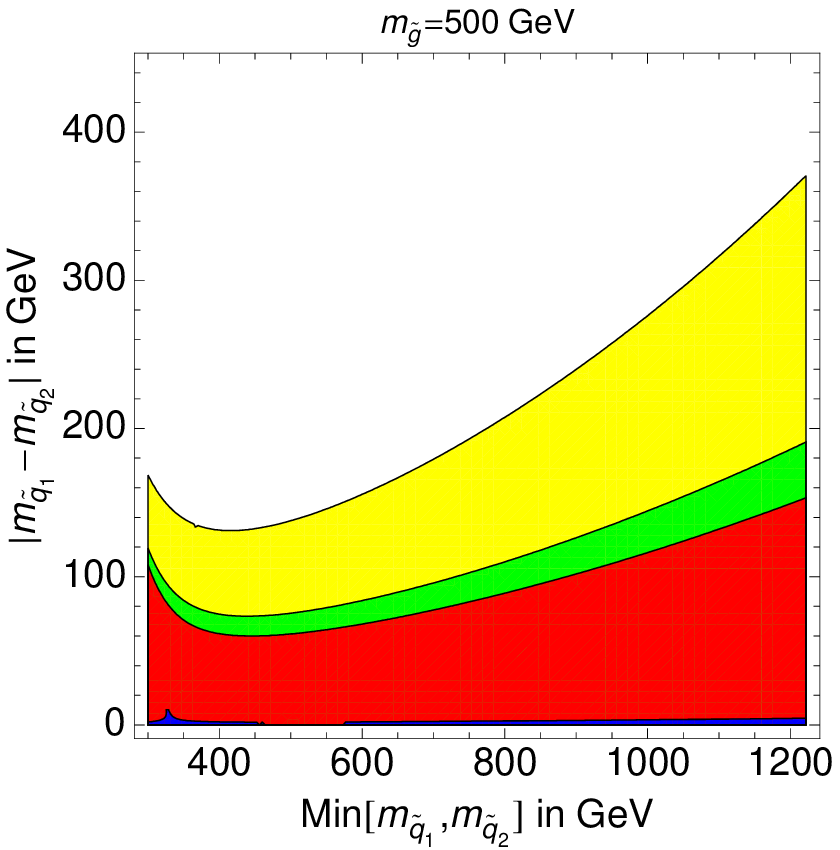}
\includegraphics[width=0.38\textwidth]{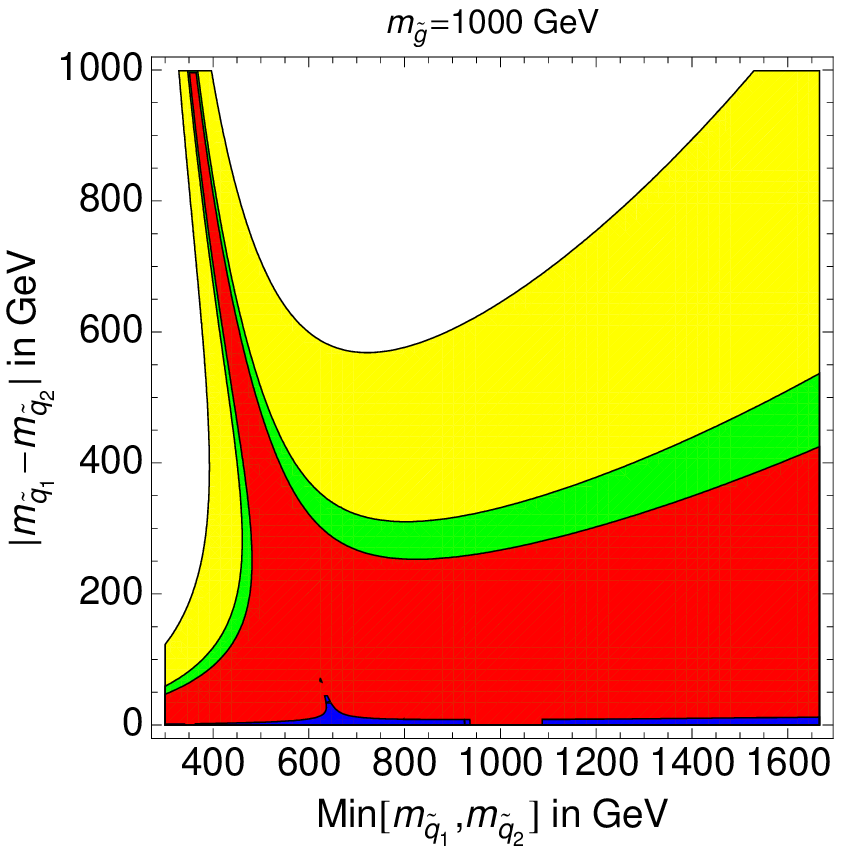}
\caption{Allowed mass splitting between the first two generations of left-handed squarks for different gluino masses. 
We assume the approximate GUT relation $M_2=(\alpha_2/ \alpha_s) m_{\tilde g}\cong 0.35$. The different colors correspond to the cases explained in the caption of Fig.~\ref{Constraints}. Note that the allowed mass splittings are large enough to permit the decay of the heavier squark into the lighter one plus a W boson.}\label{mass-splitting}
\end{figure}
\end{widetext}
While nearly all previous analyses focused on the gluino contributions to \kk\ and \dd\ mixing in the case of nonminimal flavor violation \cite{Nir:1993mx, Hagelin:1992tc, Gabbiani:1996hi, Ciuchini:1998ix, Ciuchini:2007cw, Blum:2009sk, Gedalia:2009kh}  Ref.~\cite{Altmannshofer:2009ne} included (but only numerically) the electroweak effects. However, the main focus of Ref.~\cite{Altmannshofer:2009ne} is not on the mass splitting between the squarks and the importance of the different contributions is not apparent from the scatter plots shown in their article. In our analysis we have examined in detail the size of the different contributions (neutralino, neutralino-gluino, gluino and chargino boxes) to \dd\ and \kk\ mixing in the presence of flavor off-diagonal mass-insertions in the LL sector of the squark mass matrices. It is found that gluino contributions suffer from a cancellation between the crossed and the uncrossed boxes for $m_{\tilde{g}}\approx1.5\,m_{\tilde q}$. In addition, winos couple directly to left-handed squark fields (without involving small gaugino or left-right mixing) and their contribution is not affected by such a cancellation. Therefore, we conclude that the (usually neglected) contributions from chargino, neutralino, and mixed neutralion-gluino diagrams can be of the same order as (or even dominant over) the gluino contribution especially if $M_2\approx m_{\tilde q}< m_{\tilde{g}}$.

In the analysis of the allowed mass splitting between the first two generations we focused on the "minimal case" in which the up (down) squark mass matrix is diagonal in the super-CKM basis, but not proportional to the unit matrix. In this case flavor off-diagonal elements in the down (up) sector are induced via the SU(2) relation and are therefore contribute to \kk\ (\dd) mixing. It is found that the constraints on the mass splitting are strong for light gluino masses. However, if the gluino is heavier than the squarks there are large regions in parameter space, allowed by \kk\ (\dd) mixing, with highly nondegenerate squark masses. This has interesting consequences both for LHC benchmark scenarios (which usually assume degenerate squarks for the first two generations) and for models with Abelian flavor symmetries (which predict nondegenerate squark masses for the first two generation) because \kk\ and \dd\ mixing cannot exclude non-degenerate squark masses of the first two generations.

\smallskip
{\it Acknowledgments.}~We are grateful to Ulrich Nierste for
helpful discussions and proofreading the article. We thank Sebastian Wurster for confirming our numerical results and Lars Hofer for helpful discussions. We also thank Gilad Perez for useful discussions. This work is supported by BMBF grants 05HT6VKB and 
05H09VKF and by the EU Contract No.~MRTN-CT-2006-035482, 
\lq\lq FLAVIAnet.'' Andreas Crivellin and Momchil Davidkov acknowledge the financial support of the State of
Baden-W\"urttemberg through \emph{Strukturiertes Promotionskolleg
Elementarteilchenphysik und Astroteilchenphysik}.

\bibliography{K-mixing}

\end{document}